\documentclass[pre,twocolumn]{revtex4-1}

\usepackage{graphicx}
\usepackage{dcolumn}
\usepackage{bm}

\usepackage[utf8]{inputenc}
\usepackage[T1]{fontenc}
\usepackage{mathptmx}

\usepackage{amsmath}
\usepackage{mathtools}
\usepackage{amssymb}

\usepackage{pst-node}
\usepackage{verbatim}   
\usepackage{color}      
\usepackage{subfigure}  
\usepackage{hyperref}   

\usepackage{braket}

\newcommand{\ba}{\begin{eqnarray}}
\newcommand{\ea}{\end{eqnarray}}
\newcommand{\be}{\begin{equation}}
\newcommand{\ee}{\end{equation}}

\pdfoutput=1

\begin{document}

\title{Semi-Markovian switching in a fluctuating harmonic trap: An age-structured formulation}

\author{Derek Frydel}
\affiliation{Department of Chemistry, Universidad Técnica Federico Santa María, Campus San Joaquin, Santiago, Chile}

\date{\today}

\begin{abstract}
\small{
We study a Brownian particle in a harmonic trap whose stiffness switches between two values with arbitrary waiting-time statistics, generating semi-Markovian dynamics. To treat the resulting temporal memory, we formulate the problem in an enlarged age-structured state space, restoring Markovianity and yielding a local Fokker--Planck description. Within this framework, we derive exact steady-state integral equations for the spatial and birth distributions and obtain exact expressions for stationary moments, injected power, and potential energy.
In the second part of the paper, we analyze the stochastic-resetting limit, corresponding to a particle alternately released and trapped. By representing the stationary spatial distribution as a superposition of Gaussian states with fluctuating variance, the problem can be reformulated as a switching process in variance space. This yields exact integral equations for the variance distributions and leads to a simplified description amenable to direct analytical treatment.
}
\end{abstract}

\pacs{
}

\maketitle

\section*{Introduction}

Particles confined in fluctuating external potentials have emerged as a minimal and analytically tractable setting for studying nonequilibrium dynamics.  
Because nonequilibrium systems are notoriously complex and often resist analytical treatment, models that admit exact results play an important role in revealing 
generic mechanisms and testing nonequilibrium principles.

Early work already demonstrated that temporal fluctuations of an external environment can fundamentally modify transport and first-passage behavior, as illustrated by resonant activation in fluctuating barriers~\cite{Doering-PRL-1992} and dispersion in fluctuating force fields~\cite{VandenBroeck-PRL-1994}. In recent years, interest in such systems has grown again, driven both by their conceptual simplicity and by their connections to modern nonequilibrium phenomena
~\cite{PRE-Pal-2013,JOP-Santra-2021,JOPA-Alston-2022,SoftMatt-Santra-2024a,PRE-Frydel-2024,Lowen-PRE-2024,Roldan-PRL-2024,PRE-Schehr-2024,Baziei-PRE-2025,Mukherjee-JSTAT-2025,Bello-NJP-2025}.

The present work focuses on a single Brownian particle in a harmonic trap whose stiffness fluctuates between two values. An additional motivation for studying this system is its close connection to stochastic resetting (SR), a process in which a diffusive trajectory is intermittently interrupted and restarted from a fixed initial position~\cite{PRL-Evans-2011,JPA-Evans-2013,PRE-Whitehouse-2013,JPA-Evans-2014}. A fluctuating harmonic trap approaches this behavior when the potential is intermittently switched on and off, effectively alternating between free diffusion and confinement~\cite{JOP-Santra-2021,JOPA-Alston-2022,Frydel-Chaos-2025, JOPA-Olsen-2024,PRE-Olsen-2024,Li-Chaos-2025}. This more realistic setup provides a soft-matter realization of stochastic resetting that can be explored experimentally~\cite{Nature-Trizac-2016,Tal-Friedman-2020,PRR-Besga-2020,Goerlich2024}.

Most studied scenarios assume switching occurring at constant rates, corresponding to exponential (memoryless) waiting-time distributions and therefore to fully Markovian dynamics. In realistic experimental situations, however, switching events are not necessarily restricted to exponential statistics. More general switching mechanisms naturally introduce memory and give rise to semi-Markovian dynamics.

A canonical example of semi-Markovian behavior is a particle jumping on a lattice. Even if the jumping process itself is Markovian---meaning that the next jump does not depend on the past trajectory---non-exponential waiting-time distributions generate memory. For example, jumps at equal time intervals lead to memory because the probability of a jump depends on the time already spent at the site. Only exponential waiting-time distributions generate a memoryless process.

The present system with non-exponential switching statistics is likewise semi-Markovian. The Brownian motion within each state remains Markovian, while the memory originates from the statistics of the residence times spent in each state.

The goal of this work is to develop a general theoretical framework for fluctuating harmonic potentials with semi-Markovian switching statistics. 
The only restriction imposed on the waiting-time distributions is the existence of a finite mean residence time, which ensures a well-defined stationary regime.

A convenient way to treat this class of semi-Markovian systems is to enlarge the state space by introducing variables that encode the memory of the process. 
For the present system, the relevant variable is the age, namely the time elapsed since the most recent switching event. 
This approach avoids temporally non-local formulations involving memory kernels~\cite{Hanggi-AdvChemPhys-1995,Metzler-PR-2000,Coffey-Book-2004}.

Age-structured formulations have a long history in population dynamics and stochastic processes. 
The introduction of age as an additional dynamical variable dates back to the work of McKendrick and von Foerster in population dynamics and cellular proliferation~\cite{McKendrick-1926,VonFoerster-1959}. 
Cox later showed that non-Markovian stochastic processes can be rendered Markovian through the inclusion of supplementary variables~\cite{Cox-1955-Book}. 
More recently, related age-structured approaches have been applied in statistical physics to anomalous transport, active matter, and non-Markovian dynamics~\cite{Fedotov-PRE-2007,Fedotov-PRE-2010,Angelani-NJP-2010,Farage-JCP-2014,Fedotov-PRE-2018}.


Beyond the specific fluctuating harmonic model studied here, the present framework is intended more broadly as a general approach for semi-Markovian switching dynamics with arbitrary waiting-time distributions. Such processes have recently attracted considerable interest in active matter, stochastic transport, and nonequilibrium statistical mechanics~\cite{PRE-Farago-2024,NJP-Frydel-2025,JSTAT-Santra-2023,SoftMatt-Santra-2024,Chaos-Santra-2025}.


This paper is organized as follows. In Sec.~(\ref{sec:age-formalism}), we introduce the age-structured formulation for semi-Markovian switching dynamics and derive the corresponding evolution equations in the extended state space. In Sec.~(\ref{sec:steady-state}), we develop the steady-state theory and obtain self-consistent integral equations for the stationary spatial and birth distributions. In Sec.~(\ref{sec:moments}), we derive exact second moments and use them to analyze potential energy, injected power, and representative waiting-time statistics. In Sec.~(\ref{sec:SR-limit}), we examine the stochastic-resetting limit and show that the injected power becomes universal while spatial fluctuations remain sensitive to temporal memory. 
In Sec.~(\ref{sec:gaussian-superposition}), we introduce a variance-space formulation for the stochastic-resetting limit, where the dynamics reduces to deterministic transport in variance space. 
Finally, Sec.~(\ref{sec:conclusion}) summarizes the main results and discusses possible extensions.

\section{Age-structured formalism}
\label{sec:age-formalism}

We consider a Brownian particle confined by a harmonic potential
\[
u_i(x)=\frac12 K_i x^2,
\]
whose stiffness fluctuates between two values \(K_1\) and \(K_2\).
The switching statistics are characterized by waiting-time distributions
\(r_i(a)\) (\(i\in\{1,2\}\)), which specify the residence-time density in each state.
The variable \(a\) denotes the age of the current state, defined as the time elapsed since the most recent switching event.
This age variable is distinct from the global time \(t\), which measures the total duration of the process.

Associated with each waiting-time distribution \(r_i(a)\) is the corresponding survival probability
\[
S_i(a)=\int_a^\infty da'\, r_i(a'),
\]
together with the hazard rate
\begin{equation}
R_i(a)=\frac{r_i(a)}{S_i(a)},
\label{eq:h}
\end{equation}
which represents the conditional switching rate at age \(a\), given survival in state \(i\) up to that age.

Throughout this work, we assume waiting-time distributions with finite mean residence times,
\begin{equation}
\tau_i = \int_0^\infty da\, a\,r_i(a) < \infty.
\end{equation}
Under this condition, stationary occupation probabilities and stationary switching rates are well defined.
Waiting-time distributions with divergent mean residence times do not admit a well-defined stationary regime; 
they require a separate treatment and are not considered here.

When \(r_i(a)\) is exponential,
\(
r_i(a)=\tau_i^{-1}e^{-a/\tau_i},
\)
the hazard rate becomes age independent,
\(
R_i(a)=\tau_i^{-1},
\)
and the switching dynamics is memoryless.  In this case, the dynamics in the state space \((x,i)\) is fully Markovian.

For a non-exponential waiting-time distribution, however, the hazard rate \(R_i(a)\) acquires an explicit dependence on the elapsed age \(a\). 
Consequently, the probability of switching depends not only on the current state \(i\), but also on the time already spent in that state, rendering 
the dynamics in \((x,i)\) non-Markovian.  Since the embedded switching dynamics remains Markovian while the residence-time statistics 
are non-exponential, the resulting dynamics belongs to the class of semi-Markov processes.

To emphasize this point, the non-Markovianity originates solely from the switching-time statistics and not from the spatial dynamics between switching 
events, which remains locally Brownian and memoryless. A standard way to restore Markovianity for such semi-Markov processes is to 
augment the state space with the supplementary age variable \(a\), extending the description from \((x,t,i)\) to \((x,t,i,a)\). In this enlarged 
state space, the future evolution depends only on the current values of \((x,i,a)\), yielding a fully Markovian age-structured representation.

The probability density in the extended space, \(q_i(x,t,a)\), is related to \(\rho_i(x,t)\) via
\begin{equation}
\rho_i(x,t)=\int_0^\infty da\, q_i(x,t,a).
\label{eq:rhoi}
\end{equation}
Since \(a\) has dimensions of time, the density \(q_i\) has units
\[
[q_i]=\frac{1}{\text{length}\cdot\text{time}},
\]
and therefore represents a probability density in the extended \((x,a)\) space.

The evolution of \(q_i\) in the extended space is governed by the linear Fokker--Planck equation
\begin{equation}
\partial_t q_i + \partial_a q_i
=
L_i q_i
-
R_i(a)\,q_i
+
\delta(a)\!\int_0^\infty da'\, R_j(a')\,q_j(x,t,a'),
\label{eq:age-pde-0}
\end{equation}
where the spatial operator \(L_i\) is given by
\begin{equation}
L_i=\partial_x\!\big(\mu K_i x + D\,\partial_x\big),
\label{eq:Li}
\end{equation}
and \(R_i(a)\) is defined in Eq.~(\ref{eq:h}).

In Eq.~(\ref{eq:age-pde-0}), the term \(\partial_a q_i\) represents deterministic drift in age, the term \(-R_i(a)q_i\) accounts for loss due to switching, 
and the final term describes particles entering state \(i\) at age \(a=0\). The delta function reflects the renewal structure of the process: all newly 
switched particles are reborn with zero age immediately following a transition.

Equivalently, by restricting the dynamics to \(a>0\), we eliminate the delta term, leading to
\begin{equation}
\partial_t q_i + \partial_a q_i
=
L_i q_i
-
R_i(a)q_i,
\qquad a>0,
\label{eq:age-pde}
\end{equation}
where the distribution at birth is encoded in the boundary condition
\begin{equation}
q_i(x,t,0)
=
\int_0^\infty da\, R_j(a)\,q_j(x,t,a),
\qquad j\neq i.
\label{eq:age-bc}
\end{equation}

In this formulation, only the total spatial density is normalized,
\begin{equation}
\int_{-\infty}^\infty dx\, \rho(x,t)=1,
\qquad \rho=\rho_1+\rho_2,
\label{eq:norm}
\end{equation}
while
\[
w_i(t)=\int_{-\infty}^{\infty} dx\, \rho_i(x,t)
\]
denotes the probability of finding the particle in state \(i\) at time \(t\).

Equations~(\ref{eq:age-pde})--(\ref{eq:norm}) provide a complete, local-in-time theoretical framework for the semi-Markov switching dynamics 
in the enlarged age-structured state space.

\subsubsection{Integrating out age}

The reduced description in the state space \((x,t,i)\) is recovered by integrating Eq.~(\ref{eq:age-pde}) over \(a\), yielding
\ba
\partial_t \rho_i
   &=&
   L_i\,\rho_i
   + q_i(x,t,0)
   - \int_0^\infty da\, R_i(a)\,q_i(x,t,a),
\ea
where
\[
\rho_i(x,t)=\int_0^\infty q_i(x,t,a)\,da.
\]
Using the boundary condition in Eq.~(\ref{eq:age-bc}), we obtain
\begin{equation}
\partial_t\rho_i = L_i\,\rho_i + b_i(x,t) - b_j(x,t), 
\qquad j\neq i,
\label{eq:age-ni-pde-2}
\end{equation}
where
\be
b_i(x,t)\equiv q_i(x,t,0)
\label{eq:bi}
\ee
denotes the \emph{birth flux} into state \(i\) at position \(x\) and time \(t\).
In this formulation, memory is not represented through an explicit memory kernel, but enters through the birth fluxes \(b_i(x,t)\).

For a memoryless process,
\(
r_i(a)=\tau_i^{-1}e^{-a/\tau_i},
\)
the instantaneous switching rate becomes constant,
\(
R_i(a)=\tau_i^{-1},
\)
and Eq.~(\ref{eq:age-bc}) reduces to
\[
b_i(x,t)=\frac{\rho_j(x,t)}{\tau_j}, 
\qquad j\neq i.
\]
Substituting this relation into Eq.~(\ref{eq:age-ni-pde-2}) yields
\begin{equation}
\partial_t\rho_i = L_i\,\rho_i +  \frac{\rho_j}{\tau_j} - \frac{\rho_i}{\tau_i},
\qquad j\neq i.
\end{equation}
which recovers a fully Markovian description within the reduced state space \((x,t,i)\).
For exponential waiting-time distributions \(r_i(a)\), the reduced dynamics remains Markovian.

\section{Steady-state}
\label{sec:steady-state}

Under steady-state conditions, the age-structured Fokker--Planck equation
(\ref{eq:age-pde}) reduces to
\begin{equation}
\partial_a q_i(x,a)
= L_i q_i(x,a) - R_i(a)q_i(x,a).
\label{eq:age-pde-SS}
\end{equation}
As a linear first-order partial differential equation in the age variable $a$, its general solution can be written as
\begin{equation}
q_i(x,a)
= S_i(a)\int_{-\infty}^{\infty} dx'\, G_i(x,x';a)\,b_i(x'),
\label{eq:pi-sol-1}
\end{equation}
where $G_i$ is the propagator associated with the operator $L_i$,
\[
\partial_a G_i(x,x';a)=L_i G_i(x,x';a),
\qquad
G_i(x,x';0)=\delta(x-x'),
\]
and $b_i(x)\equiv q_i(x,0)$ denotes the steady-state birth flux.
Equation~(\ref{eq:pi-sol-1}) admits a transparent physical interpretation:
particles are injected at age $a=0$ according to the distribution $b_i(x)$,
propagate under $G_i$ for a duration $a$, and survive up to that age with
probability $S_i(a)$.

Integrating Eq.~(\ref{eq:pi-sol-1}) over age yields the steady-state spatial density
\begin{equation}
\rho_i(x)
= \int_{-\infty}^\infty dx'\, b_i(x')
  \int_0^\infty da\, S_i(a)\,G_i(x,x';a).
\label{eq:pi-sol-2}
\end{equation}
To close the system, we still need a relation that defines $b_i(x)$.
Such a relation follows from the steady-state form of the boundary condition
(\ref{eq:age-bc}),
\[
b_i(x)=\int_0^\infty da\, R_j(a)q_j(x,a),
\qquad j\neq i.
\]
Substituting the solution (\ref{eq:pi-sol-1}) into this expression gives
\begin{equation}
b_i(x)
= \int_{-\infty}^{\infty} dx' \, b_j(x')
  \int_0^\infty da\, r_j(a)\,G_j(x,x';a),
  \qquad j\neq i, 
\label{eq:bi-sol-2}
\end{equation}
where we used the identity $S_j(a)R_j(a)=r_j(a)$.
Equations~(\ref{eq:pi-sol-2}) and (\ref{eq:bi-sol-2}) therefore form a closed,
self-consistent set determining both the spatial densities $\rho_i(x)$ and
the birth fluxes $b_i(x)$.

The two equations are structurally similar and differ only in their age weighting.
In Eq.~(\ref{eq:pi-sol-2}), trajectories contribute continuously at all ages until
they terminate, whereas in Eq.~(\ref{eq:bi-sol-2}) they contribute only at the instant
of switching. This distinction highlights the role of the birth flux as the sole carrier
of memory in the reduced description.

Integrating Eqs.~(\ref{eq:pi-sol-2}) and (\ref{eq:bi-sol-2}), together with the normalization condition
\(
\int_{-\infty}^\infty dx\,(\rho_1+\rho_2)=1,
\)
yields
\[
\int_{-\infty}^\infty dx\, b_i(x) = \frac{1}{\tau_1+\tau_2},
\qquad
\int_{-\infty}^\infty dx\, \rho_i(x) = \frac{\tau_i}{\tau_1+\tau_2},
\]
where
\(
\tau_i = \int_0^\infty da\, a\,r_i(a) = \int_0^\infty da\, S_i(a)
\)
is the mean residence time in state \(i\). 
As expected, the occupation probability of each state is proportional to the corresponding mean residence time, providing a consistency check of the formalism.

For completeness, we give the explicit form of the propagator corresponding to Ornstein--Uhlenbeck dynamics,
\begin{equation}
G_i(x,x';a) = \frac{1}{\sqrt{2\pi \Sigma_i}}
\exp\!\left[-\frac{(x - m_i x')^2}{2\Sigma_i}\right],
\label{eq:Gxx}
\end{equation}
with mean and variance
\begin{equation}
m_i x' = x' e^{-\mu K_i a},
\qquad
\Sigma_i = \frac{D}{\mu K_i} \left(1 - e^{-2\mu K_i a}\right).
\label{eq:Sigma}
\end{equation}

\subsection{Normalized formulation}

The relations~(\ref{eq:pi-sol-2}) and~(\ref{eq:bi-sol-2}) provide a complete steady--state description 
of the fluctuating harmonic potential. It is convenient to recast these equations in a normalized form 
by introducing the distributions \(n_i(x)\) and \(n_{b_i}(x)\), defined as
\[
\rho_i(x)=\frac{\tau_i}{\tau_1+\tau_2}\,n_i(x),
\qquad
b_i(x)=\frac{1}{\tau_1+\tau_2}\,n_{b_i}(x).
\]

The corresponding normalization conditions
\[
\int_{-\infty}^{\infty} dx\, n_i(x)=1,
\qquad
\int_{-\infty}^{\infty} dx\, n_{b_i}(x)=1
\]
follow directly from the definitions above together with
\[
\int_0^\infty da\, S_i(a)=\tau_i.
\]

In terms of these normalized distributions, Eqs.~(\ref{eq:pi-sol-2}) and~(\ref{eq:bi-sol-2}) become
\begin{align}
n_{b_1}(x)
 &= \int_{-\infty}^{\infty} dx'\, n_{b_2}(x')\!\left[\int_0^\infty da\, r_2(a)\,G_2(x,x';a)\right], \nonumber\\
n_{b_2}(x)
 &= \int_{-\infty}^{\infty} dx'\, n_{b_1}(x')\!\left[\int_0^\infty da\, r_1(a)\,G_1(x,x';a)\right],
\label{eq:b1-sol-2B}
\end{align}
and
\begin{align}
n_1(x)
 &= \frac{1}{\tau_1}\int_{-\infty}^{\infty} dx'\, n_{b_1}(x')\!\left[\int_0^\infty da\, S_1(a)\,G_1(x,x';a)\right], \nonumber\\
n_2(x)
 &= \frac{1}{\tau_2}\int_{-\infty}^{\infty} dx'\, n_{b_2}(x')\!\left[\int_0^\infty da\, S_2(a)\,G_2(x,x';a)\right].
\label{eq:pi-sol-2B}
\end{align}

The four coupled integral equations~(\ref{eq:b1-sol-2B})--(\ref{eq:pi-sol-2B}) form the basis
for the steady--state analysis of the fluctuating harmonic potential.
The equations for the birth distributions \(n_{b_i}(x)\) form a closed, self-consistent system.
Once these distributions are known, the stationary spatial densities \(n_i(x)\)
follow directly from Eqs.~(\ref{eq:pi-sol-2B}).

In the memoryless limit, we have
\(
r_i(a)=\frac{1}{\tau_i}S_i(a),
\)
and Eqs.~(\ref{eq:b1-sol-2B}) and~(\ref{eq:pi-sol-2B}) imply
\[
n_1(x)=n_{b_2}(x),
\qquad
n_2(x)=n_{b_1}(x).
\]
The four coupled integral equations therefore reduce to two coupled equations for the stationary spatial densities, recovering the standard Markovian description.

This result admits a transparent physical interpretation. In the memoryless case, the stationary spatial distribution in one state coincides with the distribution 
of particle positions immediately prior to a transition, which then becomes the birth distribution of the opposing state. Consequently, the stationary state may 
be reconstructed entirely by sampling only the endpoints of the trajectories. 
This equivalence no longer holds for non-exponential waiting-time statistics, where the age structure contributes explicitly to the stationary spatial distribution.

A similar structure appears in run-and-tumble dynamics with exponentially distributed run times, where the stationary density coincides with the distribution of tumbling events. 
This equivalence permits an alternative description of the process in terms of a jump process and associated renewal integral equations 
\cite{NJP-Frydel-2025,POF-Frydel-2024}.

\section{Second moments}
\label{sec:moments}

Although the integral equations in Eqs.~(\ref{eq:b1-sol-2B}--\ref{eq:pi-sol-2B}) are generally difficult to solve explicitly, 
the formalism nonetheless provides a structure that can be used to derive exact expressions for simpler observables. 
We are specifically interested in the second moments of the stationary distributions.

To obtain these moments, we operate on Eqs.~(\ref{eq:b1-sol-2B}--\ref{eq:pi-sol-2B}) with the integral operator
\(
\int_{-\infty}^{\infty} dx\, x^2 .
\)
Then, using
\be
\int_{-\infty}^{\infty} x^2\,G_i(x,x';a)\,dx = \Sigma_i(a)+m_i(a)^2x'^2,
\label{eq:x2G}
\ee
where \(m_i(a)x'\) and \(\Sigma_i(a)\) are the mean and variance of the propagator defined in Eq.~(\ref{eq:Sigma}), we arrive 
at the following formulas:
\ba
\langle x^2\rangle_{b_1} &=& \frac{k_BT}{K_1} \left[ 1-\left(1-\frac{K_1}{K_2}\right)\frac{A}{\eta_1} \right],
\nonumber\\
\langle x^2\rangle_{b_2} &=& \frac{k_BT}{K_2} \left[ 1-\left(1-\frac{K_2}{K_1}\right)\frac{A}{\eta_2} \right],
\label{eq:x2b}
\ea
and
\ba
\langle x^2\rangle_{1} &=& \frac{k_BT}{K_1} \left[ 1-\left(1-\frac{K_1}{K_2}\right) \frac{A\tau_{K_1}}{2\tau_1} \right],
\nonumber\\
\langle x^2\rangle_{2} &=& \frac{k_BT}{K_2} \left[ 1-\left(1-\frac{K_2}{K_1}\right) \frac{A\tau_{K_2}}{2\tau_2} \right].
\label{eq:x2}
\ea

Here, the averages \(\langle x^2\rangle_{b_i}\) and \(\langle x^2\rangle_i\) refer to the birth and spatial distributions, respectively. 
To simplify the results, we introduce the relaxation time
\[
\tau_{K_i} = \frac{1}{\mu K_i},
\]
which characterizes the return to equilibrium in the harmonic trap. 
We also define the dimensionless parameter
\be
\eta_i = 1 - \tilde r_i\!\left(\frac{2}{\tau_{K_i}}\right),
\label{eq:eta-def}
\ee
where \( \tilde r_i(s) = \int_0^{\infty} da \, e^{-sa} r_i(a) \) is the Laplace transform of the waiting-time distribution. Finally,
\be
A = \frac{\eta_1\eta_2} {\eta_1+\eta_2-\eta_1\eta_2}
\label{eq:A}
\ee
is a monotonically increasing function of both \(\eta_1\) and \(\eta_2\).  Since the parameters \(\eta_i\) depend on the full structure 
of the waiting-time distributions \(r_i(a)\), and not only on their mean residence times, they encode the memory effects associated 
with the semi-Markovian dynamics. Only for exponential waiting-time distributions can \(\eta_i\) be expressed solely in terms of the 
mean waiting times,
\be
\eta_i = \frac{\tau_i} {\tau_i + \tau_{K_i}/2}, \qquad r_i(a)=\tau_i^{-1}e^{-a/\tau_i}.  
\label{eq:eta-exp}
\ee

To verify these expressions, we check that they recover the expected behavior in several limiting cases. 
When the trap fluctuations are suppressed, \(K_1=K_2\), the system should reduce to equilibrium dynamics. 
Indeed, the second (coupling) terms in Eqs.~(\ref{eq:x2b}) and (\ref{eq:x2}) vanish in this limit, yielding
\be
\langle x^2\rangle_{b_i} = \langle x^2\rangle_i = \frac{k_BT}{K_i},
\ee
which is the standard equilibrium result.

Another important limit is the Markovian case, where the waiting-time distributions are exponential. 
Substituting Eq.~(\ref{eq:eta-exp}) into Eq.~(\ref{eq:x2}) yields
\[
\langle x^2\rangle_i = \langle x^2\rangle_{b_j}
=
\frac{k_BT}{K_i}
\left[
1+
\frac{\tau_j(\tau_{K_i}-\tau_{K_j})}
{\tau_{K_j}(\tau_1+\tau_2)+2\tau_1\tau_2}
\right]^{-1},
\qquad i\neq j,
\]

Finally, we consider the symmetric fast-switching regime
\[
\tau_1 = \tau_2\equiv\tau,
\qquad
\tau\to0.
\]
In this limit, the residence times become much shorter than the relaxation timescales of the harmonic traps.
Expanding the exponential in Eq.~(\ref{eq:eta-def}) in the fast-switching limit, where the mean residence times satisfy \(\tau\ll\tau_{K_i}\), gives
\[
\eta_i = 1-\int_0^\infty da\, r_i(a)e^{-2a/\tau_{K_i}}
\simeq
\frac{2\tau}{\tau_{K_i}},
\]
so that \(\eta_i\ll1\). Consequently,
\[
A = \frac{\eta_1\eta_2}{\eta_1 + \eta_2 - \eta_1 \eta_2}
\simeq
\frac{2\tau}{\tau_{K_1} + \tau_{K_2}}.
\]
Substituting these expressions into Eqs.~(\ref{eq:x2b})--(\ref{eq:x2}) yields
\be
\langle x^2\rangle_{b_i} = \langle x^2\rangle_i = \frac{2k_BT}{K_1+K_2}.
\label{eq:x2-Keff}
\ee
Thus, in the fast-switching limit, the system behaves as an effective harmonic trap with stiffness
\[
K_{\rm eff}=\frac{K_1+K_2}{2}.
\]
For exponentially distributed waiting times, this effective-medium limit was previously obtained by Santra \emph{et al.}~\cite{JOP-Santra-2021} from the exact stationary distribution. 
The present derivation shows that the same effective stiffness emerges more generally for arbitrary waiting-time distributions, provided the mean residence times remain short 
compared to the relaxation times of the individual traps.

Note the equilibrium-like form of Eq.~(\ref{eq:x2-Keff}) does not imply true equilibrium. Maintaining the fast-switching regime requires continuous work injection, leading to persistent heat dissipation and entropy production. For the symmetric case $\tau_1=\tau_2\equiv\tau$, this regime in fact corresponds to maximal entropy production as a function of $\tau$~\cite{PRE-Frydel-2022,PRE-Frydel-2023a}. We will show in the subsequent sections that the injected power remains finite in this limit.


\subsection{Potential energy}

The average potential energy of the fluctuating harmonic trap is
\be
PE = \frac{K_1}{2}\,\frac{\tau_1}{\tau_1+\tau_2}\,\langle x^2\rangle_1 + \frac{K_2}{2}\,\frac{\tau_2}{\tau_1+\tau_2}\,\langle x^2\rangle_2,
\label{eq:PE-gen}
\ee
where \(\tau_i/(\tau_1+\tau_2)\) is the probability of finding the particle in state \(i\).
Substituting Eq.~(\ref{eq:x2}) leads to extensive cancellations and yields
\be
PE=\frac{k_BT}{2}.
\label{eq:PE-harm}
\ee
Thus, $PE$ is independent of both the trap stiffnesses and the switching statistics, 
and coincides with the equilibrium value of a static harmonic trap.

This result follows from a generalized virial relation.
Consider a Brownian particle evolving in an arbitrary time-dependent potential \(u(x,t)\),
\[
\partial_t \rho
=
\mu\,\partial_x\!\big[(\partial_x u)\rho\big]
+
D\,\partial_{xx}\rho.
\]
Multiplying by \(x^2\) and integrating over \(x\) yields
\[
\frac{d}{dt}\langle x^2\rangle
=
-2\mu\,\langle x\,\partial_x u\rangle
+
2D.
\]
Using \(D=\mu k_BT\), stationarity implies
\be
\langle x\,\partial_x u\rangle = k_BT.
\label{eq:virial}
\ee
For homogeneous potentials
\be
u(x,t)=\frac{K(t)}{p}x^p,
\label{eq:Up}
\ee
Eq.~(\ref{eq:virial}) gives
\be
PE = \frac{1}{p}\langle K(t)x^p\rangle = \frac{k_BT}{p}.
\label{eq:PE-p}
\ee
Thus, any fluctuating potential of the form \(u(x,t)=K(t)x^p/p\) for which the moment \(\langle K(t)x^p\rangle\) reaches 
a stationary value possesses the same average potential energy as its equilibrium counterpart.
For the harmonic case (\(p=2\)), Eq.~(\ref{eq:PE-p}) reduces to Eq.~(\ref{eq:PE-harm}).

The key finding of this section is that the average potential energy is entirely independent of the switching statistics and associated memory effects. 
While the individual spatial moments \(\langle x^2 \rangle_i\) remain sensitive to the semi-Markovian nature of the process, their weighted combination 
is constrained by the generalized virial relation.

\subsection{Work}

Another physical observable accessible from the second moments is the rate of work injected into the system by fluctuations of the external potential.
For a time-dependent potential \(u(x,t)\), the instantaneous work injection rate is
\be
\dot W = \frac{\partial u(x,t)}{\partial t}.
\label{eq:dW}
\ee
For the fluctuating harmonic potential
\be
u(x,t)=\frac{1}{2}K(t)x^2,
\ee
this becomes
\be
\dot W
=
\frac{1}{2}\,\dot K(t)\,x^2.
\label{eq:dW2}
\ee

In the two-state model, \(K(t)\) switches discontinuously between \(K_1\) and \(K_2\), so that \(\dot K(t)\) consists of delta spikes 
localized at switching events. Consequently, Eq.~(\ref{eq:dW2}) samples \(x^2\) only at state transitions, implying that the relevant 
moments are those of the birth distributions.

Averaging over many switching cycles yields
\be
\langle \dot W \rangle = \frac{K_1-K_2}{2}\, \frac{\langle x^2\rangle_{b_1}-\langle x^2\rangle_{b_2}} {\tau_1+\tau_2},
\label{eq:Wdot-harm}
\ee
where \(1/(\tau_1+\tau_2)\) is the average frequency of a complete switching cycle.  
The injected power is therefore controlled by the difference between the second moments of the birth distributions.

Using the explicit expressions for the moments in Eq.~(\ref{eq:x2b}) yields
\be
\langle \dot W \rangle = \frac{k_BT}{2}\, \frac{(K_1-K_2)^2}{K_1K_2}\, \frac{A}{\tau_1+\tau_2},
\label{eq:dW-final}
\ee
where \(A\) is defined in Eq.~(\ref{eq:A}). 
The parameter \(A\) satisfies \(0\le A\le 1\). 
It is the only factor in Eq.~(\ref{eq:dW-final}) that depends on the detailed form of the waiting-time distributions beyond their mean residence times.

In the steady state, the average internal energy is constant. 
Therefore the injected work must be balanced by heat dissipated into the bath,
\be
\langle \dot Q\rangle = \langle \dot W\rangle .
\ee
Thus Eq.~(\ref{eq:dW-final}) also gives the steady-state heat dissipation rate. 
Equivalently, the entropy production rate is
\[
\Pi=\frac{\langle \dot W\rangle}{T}.
\]

It is clear from Eq.~(\ref{eq:dW-final}) that at equilibrium, \(K_1=K_2\), the injected power vanishes,
\(
\langle \dot W\rangle = 0,
\)
consistent with true equilibrium.
Another interesting case is the symmetric fast-switching regime,
\[
\tau_1=\tau_2\equiv\tau,
\qquad
\tau\to0.
\]
According to Eq.~(\ref{eq:x2-Keff}), the second moments of the stationary distributions take the equilibrium-like form with effective stiffness
\(
K_{\rm eff}=\frac{K_1+K_2}{2}.
\)
In this limit, Eq.~(\ref{eq:dW-final}) reduces to
\be
\langle \dot W \rangle
=
\frac{D}{2}\,
\frac{(K_1-K_2)^2}{K_1+K_2},
\label{eq:dW-tau-0}
\ee
showing that the system remains fundamentally nonequilibrium despite the equilibrium-like form of the stationary second moments. 
Maintaining the fast-switching regime requires continuous work injection.

\subsection{Representative waiting-time distributions}
\label{subsec:r}

For fixed mean residence times \(\tau_i\), the injected power in
Eq.~\eqref{eq:dW-final} depends on the waiting-time distributions only through
the parameters \(\eta_i\), or equivalently through the combination
\(A(\eta_1,\eta_2)\) defined in Eq.~\eqref{eq:A}. Since
\[
\frac{\partial A}{\partial \eta_i}>0,
\qquad 0<\eta_i<1,
\]
maximizing the injected power is equivalent to maximizing each \(\eta_i\).

Using Eq.~\eqref{eq:eta-def}, we may write
\[
\eta_i
=
1-\left\langle e^{-2a/\tau_{K_i}}\right\rangle_{r_i}.
\]
Thus, for fixed mean \(\tau_i\), the problem reduces to minimizing
\(
\langle e^{-2a/\tau_{K_i}} \rangle_{r_i}.
\)
Since \(e^{-2a/\tau_{K_i}}\) is a convex function of \(a\), Jensen's inequality gives
\[
\left\langle e^{-2a/\tau_{K_i}}\right\rangle_{r_i}
\ge
e^{-2\tau_i/\tau_{K_i}}.
\]
Consequently,
\[
\eta_i
\le
1-e^{-2\tau_i/\tau_{K_i}}.
\]
This establishes an upper bound on the injected power.
The bound is attained exactly by the deterministic waiting-time distribution
\begin{equation}
r_i(a)=\delta(a-\tau_i),
\label{eq:single}
\end{equation}
corresponding to perfectly regular switching. Therefore, for fixed mean
residence times, deterministic switching maximizes the injected power.

Although the lower bound \(\eta_i\to0\) cannot be reached exactly by an
admissible distribution with finite mean \(\tau_i\), it can be approached
arbitrarily closely by distributions characterized by many very short residence
times together with rare long trapping events.
To illustrate how different forms of temporal memory affect the injected power,
and how they approach the limit \(\eta_i\to0\), 
we consider three representative families of waiting-time distributions.

\paragraph*{Power-law distribution.}

We first consider a heavy-tailed waiting-time distribution,
\begin{equation}
r_i(a)
=
\frac{1}{\tau_i}
\frac{k-1}{k-2}
\left[
1+\frac{a}{\tau_i(k-2)}
\right]^{-k},
\qquad k>2.
\label{eq:powerlaw-ri}
\end{equation}
which decays algebraically as
\[
r_i(a)\sim a^{-k}.
\]
The restriction \(k>2\) ensures finite mean residence times.
As \(k\to2\) from above, the residence times exhibit strong fluctuations and
rare long events become increasingly important. In the opposite limit
\(k\to\infty\), the distribution becomes progressively localized around its mean value.

The exact analytical expression for $\eta_i$ is 
\[
\eta_i = 1-(k-1)e^{z_i}z_i^{k-1}\Gamma(1-k,z_i)
\]
where 
$
z_i = {2\tau_i(k-2)} / {\tau_{K_i}}.
$

\paragraph*{Gamma distribution.}

A second example is the Gamma distribution,
\begin{equation}
r_i(a)
=
\frac{1}{\tau_i}
\frac{k}{\Gamma(k)}
\left(
\frac{k a}{\tau_i}
\right)^{k-1}
e^{-k a/\tau_i},
\qquad k\ge1.
\label{eq:gamma-ri}
\end{equation}
which naturally interpolates between Markovian and deterministic switching.
For \(k=1\), Eq.~\eqref{eq:gamma-ri} reduces to the exponential waiting-time distribution corresponding to memoryless switching.
Increasing \(k\) progressively narrows the distribution, and in the limit \(k\to\infty\) the deterministic
distribution \(r_i(a)=\delta(a-\tau_i)\) is recovered.

The parameter $\eta_i$ in this case is
$$
\eta_i = 1 -  \left( 1+\frac{2\tau_i}{k\tau_{K_i}} \right)^{-k}.  
$$

\paragraph*{Two-point distribution.}

Finally, we consider the discrete distribution
\begin{equation}
r_i(a) = \frac{1}{k}\delta(a) + \left(1-\frac1k\right) \delta\!\left( a-\frac{k\tau_i}{k-1} \right), \qquad k>1.
\label{eq:twopoint-ri}
\end{equation}
which also possesses mean residence time \(\tau_i\).  For \(k\to\infty\), the distribution approaches deterministic switching, 
and as \(k\to1^+\), the distribution approaches a strongly intermittent regime: 
most transitions occur almost instantaneously, while rare events
produce exceptionally long residence times.

The parameter $\eta_i$ in this case is
\[
\eta_i = \left(1-\frac1k\right) \left[ 1 - \exp\!\left( - \frac{k}{k-1}  \frac{2\tau_i}{\tau_{K_i}} \right) \right].
\]

\begin{figure}[t]
\begin{center}
\includegraphics[width=0.42\textwidth]{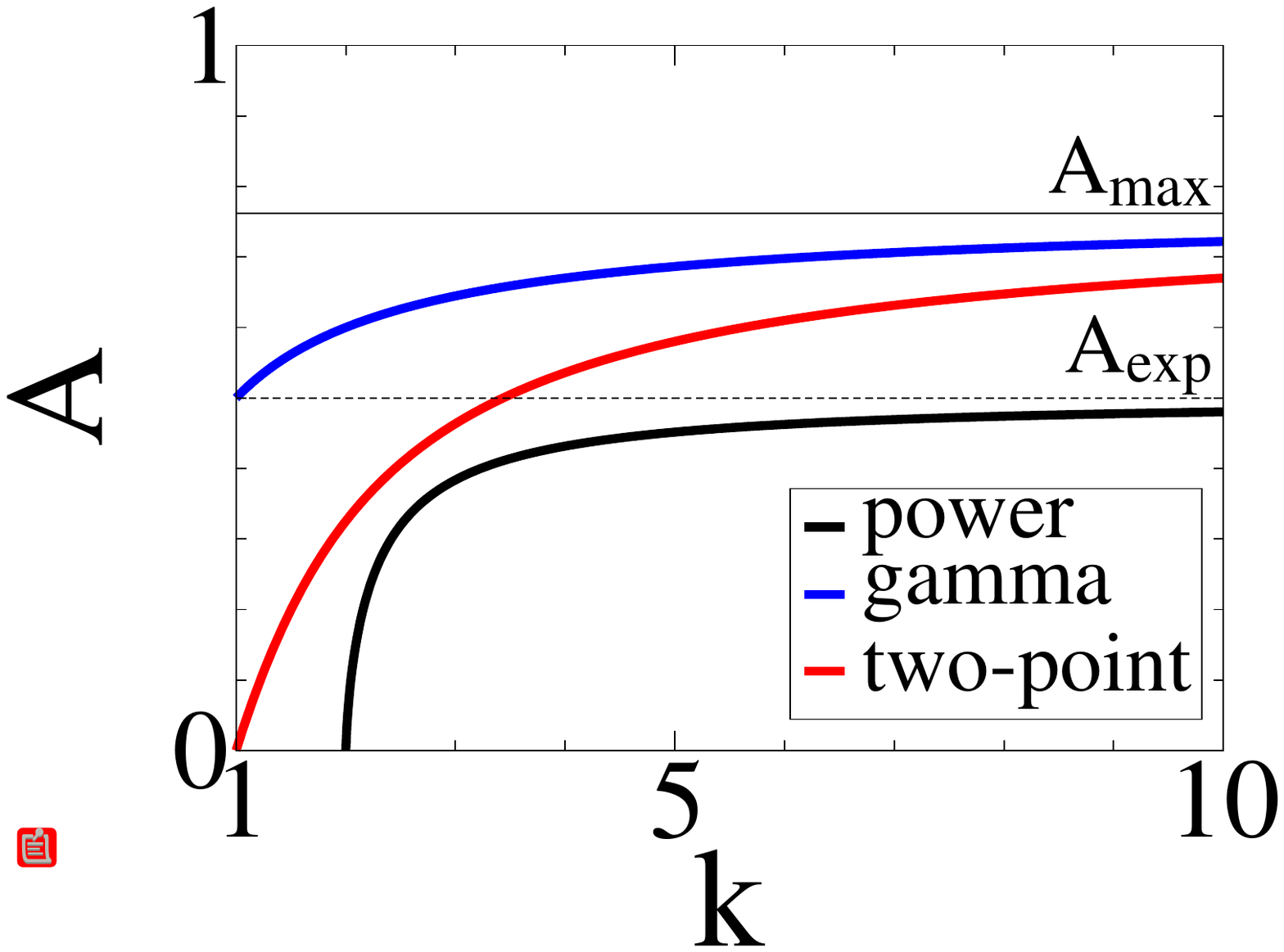}
\end{center}
\caption{
Parameter \(A\) as a function of the shape parameter \(k\)
for the Gamma, power-law, and intermittent two-point
waiting-time distributions.
The horizontal dashed lines correspond to the deterministic upper bound
\(A_{\max}=\tanh(1)\approx0.7616\)
and the memoryless (exponential) value
\(A_{\rm exp}=0.5\).
Parameters:
\(
\tau_{K_1}=\tau_{K_2}=\tau_1=\tau_2=1.
\)
}
\label{fig:Ak}
\end{figure}

To compare the influence of different waiting-time statistics,
Fig.~\ref{fig:Ak} shows the parameter \(A\) as a function of \(k\)
for fixed \(\tau_i\) and \(K_i\). Since the injected power in
Eq.~\eqref{eq:dW-final} is proportional to \(A\), the same behavior
directly determines the dependence of \(\langle \dot W\rangle\) on \(k\).
\graphicspath{{figures/}}

The Gamma distribution interpolates between the memoryless
(exponential) and deterministic switching limits.
However, regardless of \(k\), it never reaches the
\(A=0\) (zero injected power) limit.
By contrast, the power-law distribution strongly suppresses \(A\)
as the algebraic tail \(a^{-k}\) approaches the limit \(k\to2^+\).
Only the intermittent two-point distribution spans the entire
admissible interval \(0<A\le A_{\max}\).

Since the memoryless process lies within the interior of the interval
\(0<A\le A_{\max}\), temporal memory can either enhance or suppress
\(A\), and therefore the injected power
\(\langle \dot W\rangle\).

The suppression of \(A\) in the intermittent regime has a clear physical
origin. When transitions occur almost instantaneously, the particle has
insufficient time to respond to the change of the external potential.
Conversely, the rare long residence times generated by the heavy tail allow
the particle to nearly equilibrate within a given state before the next
transition occurs. Although such events are statistically rare, they persist
for exceptionally long times and therefore dominate the stationary ensemble
of trajectories. In both cases, the birth moments
\(\langle x^2\rangle_{b_1}\) and
\(\langle x^2\rangle_{b_2}\)
approach one another, causing the injected power
\(
\langle \dot W\rangle
\)
to become progressively suppressed.

\section{Stochastic resetting limit}
\label{sec:SR-limit}

As discussed in the introduction, fluctuating harmonic traps provide a natural physical realization of stochastic resetting (SR), 
in which a particle alternates between intervals of free diffusion and a trapped state.  
Within the present framework, the SR-like limit is obtained by taking
\[
K_1\to0,
\qquad
K_2=K.
\]
Taking the limit \(K_1\to0\), while keeping all other parameters finite, gives
\[
\frac{A}{\eta_1}
=
1 - 2\mu K_1\tau_1
\left(
\frac{1}{\eta_2}-1
\right)
+O(K_1^2).
\]
Identifying state \(1\) with the released state and state \(2\) with the confined state,
\[
\tau_1=\tau_{\rm off},
\qquad
\tau_2=\tau_{\rm on},
\qquad
\eta_2=\eta_{\rm on},
\]
and substituting into Eq.~(\ref{eq:x2b}), we obtain
\ba
\langle x^2\rangle_{b,\mathrm{off}} &=&
\frac{k_BT}{K}
+
2D\,\tau_{\mathrm{off}}
\left(
\frac{1}{\eta_{\mathrm{on}}}-1
\right),
\nonumber\\
\langle x^2\rangle_{b,\mathrm{on}} &=&
\frac{k_BT}{K}
+
\frac{2D\,\tau_{\mathrm{off}}}{\eta_{\mathrm{on}}},
\label{eq:x2b-SR}
\ea
where
\[
\eta_{\mathrm{on}}
=
1-\int_0^\infty da\, r_{\mathrm{on}}(a)e^{-2a/\tau_K},
\qquad
\tau_K=\frac{1}{\mu K}.
\]
Here \(\tau_{\rm off}\) and \(\tau_{\rm on}\) denote the mean residence times in the released and confined states, respectively.

Note that if the confined residence time is short compared with the trap relaxation time,
\(\tau_{\rm on}\ll \tau_K\), then
\[
\eta_{\rm on}\simeq \frac{2\tau_{\rm on}}{\tau_K}.
\]
In this regime, Eq.~(\ref{eq:x2b-SR}) gives
\ba
\langle x^2\rangle_{b,\mathrm{off}}
&\simeq&
\frac{k_BT}{K}
+
D\,\tau_{\mathrm{off}}
\left(
\frac{\tau_K}{\tau_{\mathrm{on}}}-2
\right),
\nonumber\\
\langle x^2\rangle_{b,\mathrm{on}}
&\simeq&
\frac{k_BT}{K}
+
D\,\tau_{\mathrm{off}}
\frac{\tau_K}{\tau_{\mathrm{on}}}.
\nonumber
\ea
Thus, when the trapped phase is too short to substantially restore the particle toward the origin, the birth moments grow as
\[
\langle x^2\rangle_b
\sim
\frac{k_BT}{K}
\frac{\tau_{\rm off}}{\tau_{\rm on}},
\]
indicating progressively weaker localization by the resetting mechanism. In this regime,
\(
\langle x^2\rangle_{b,\mathrm{off}}
\simeq
\langle x^2\rangle_{b,\mathrm{on}},
\)
showing that the two dynamical states become statistically indistinguishable despite the ongoing switching dynamics.

Having obtained the second moments in Eq.~(\ref{eq:x2b}), we are now in a position to evaluate the injected power using Eq.~(\ref{eq:Wdot-harm}), 
yielding
\begin{equation}
\langle \dot W \rangle
=
DK\,
\frac{\tau_{\mathrm{off}}}
{\tau_{\mathrm{off}}+\tau_{\mathrm{on}}}.
\label{eq:dW-0}
\end{equation}
Unlike the general result in Eq.~(\ref{eq:dW-final}), the injected power in the stochastic-resetting limit depends only on the mean 
residence times and is otherwise completely independent of the waiting-time statistics. Thus, while temporal memory remains encoded 
in the spatial fluctuations through \(\eta_{\rm on}\), the energetic cost of maintaining the resetting state is universal and depends only 
on the average fraction of time spent in the released state.

\begin{figure}[t]
\begin{center}
\includegraphics[width=0.25\textwidth]{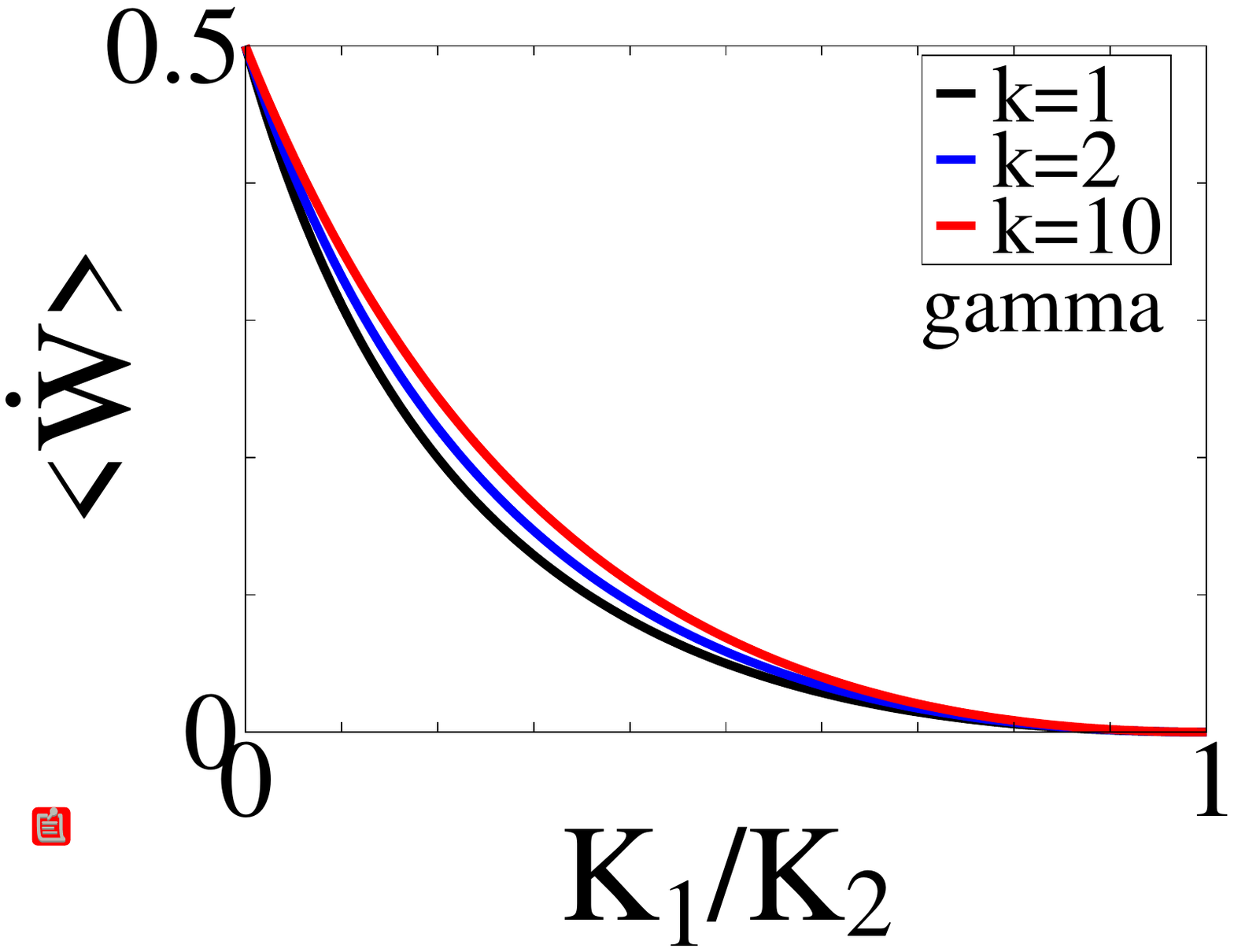}\\[0.2cm]
\includegraphics[width=0.25\textwidth]{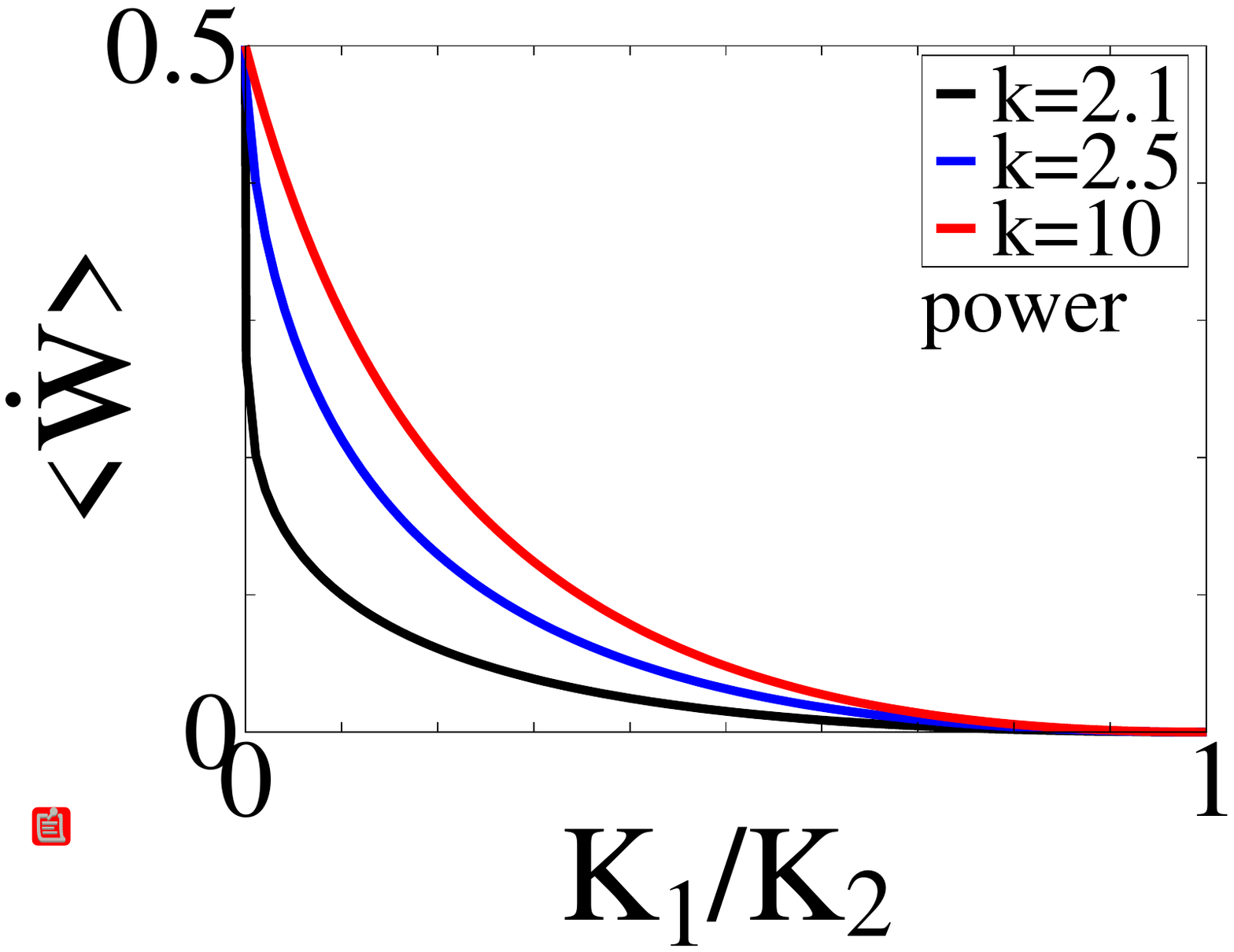}\\[0.2cm]
\includegraphics[width=0.25\textwidth]{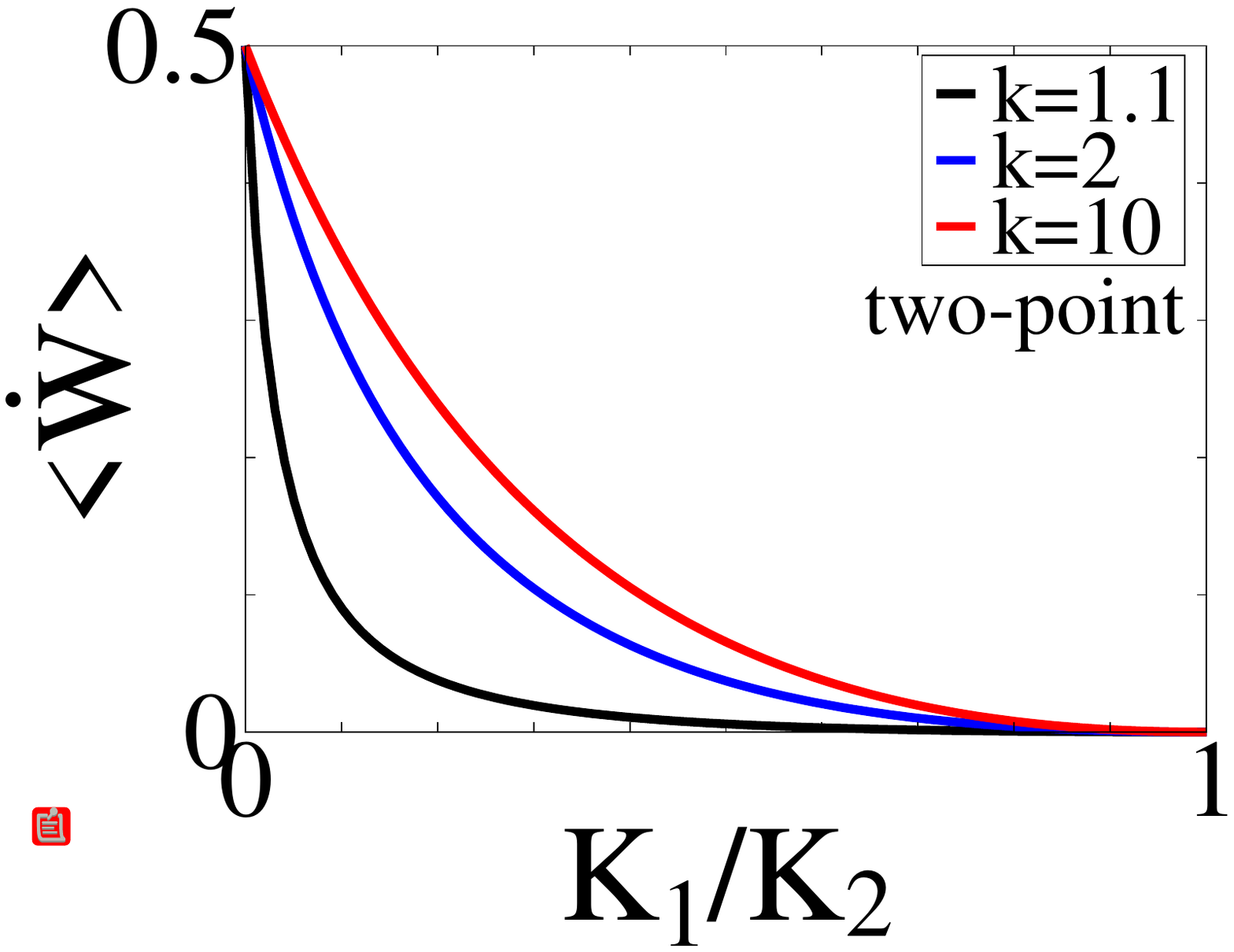}
\end{center}
\caption{
Injected power \(\langle \dot W\rangle\) as a function of \(K_1\)
for the Gamma, power-law, and two-point waiting-time distributions.
As \(K_1\to0\), all curves converge to the universal stochastic-resetting limit
\(\langle \dot W\rangle = DK\,\tau_{\rm off}/(\tau_{\rm off}+\tau_{\rm on}) \).
Parameters: \(\mu K_2=\tau_{\rm off}=\tau_{\rm on}=k_BT=1\).
}
\label{fig:WK1}
\end{figure}

To illustrate how this universality emerges in the limit \(K_1\to0\), Fig.~(\ref{fig:WK1}) shows the injected power
\(\langle \dot W \rangle\)
as a function of \(K_1\) over the interval
\(
0\le K_1\le K_2.
\)
The curves are evaluated for the different waiting-time distributions introduced in Sec.~(\ref{subsec:r}) and for several values of the parameter \(k\).
\graphicspath{{figures/}}
All curves intersect at \(K_1=K_2\), corresponding to equilibrium,
\(
\langle \dot W \rangle = 0,
\)
and again at \(K_1=0\), corresponding to the stochastic-resetting limit, where the injected power becomes 
independent of the detailed form of the waiting-time distributions, as predicted by Eq.~(\ref{eq:dW-0}).

We next derive expressions for the stationary second moments
\(
\langle x^2 \rangle_{\mathrm{off}}
\)
and
\(
\langle x^2 \rangle_{\mathrm{on}}.
\)
In the stochastic-resetting limit, Eq.~(\ref{eq:x2}) yields
\ba
\langle x^2 \rangle_{\mathrm{off}} &=& \langle x^2 \rangle_{b,\mathrm{off}} + \frac{D}{\tau_{\mathrm{off}}}\, \langle a^2\rangle_{\mathrm{off}},
\nonumber\\
\langle x^2 \rangle_{\mathrm{on}} &=& \frac{k_BT}{K} \left( \frac{\tau_{\mathrm{off}}+\tau_{\mathrm{on}}} {\tau_{\mathrm{on}}} \right),
\label{eq:x2-SR}
\ea
where
\(
\langle a^2\rangle_{\mathrm{off}} = \int_0^\infty da\, a^2\, r_{\mathrm{off}}(a),
\)
is the second moment of the released-state waiting-time distribution.

The moment
\(
\langle x^2 \rangle_{\mathrm{on}}
\)
depends only on the mean residence times.
By contrast,
\(
\langle x^2 \rangle_{\mathrm{off}}
\)
retains explicit sensitivity to the detailed form of the waiting-time distributions.
It depends on \(r_{\mathrm{on}}\) through \(\eta_{\mathrm{on}}\) in
\(
\langle x^2 \rangle_{b,\mathrm{off}}
\),
and on the second moment
\(
\langle a^2\rangle_{\mathrm{off}}
\)
of the released-state waiting-time distribution.

As a consequence, heavy-tailed waiting-time statistics in the released state can produce divergent spatial fluctuations. For example, if
\[
r_{\mathrm{off}}(a)\sim a^{-k},
\qquad
2< k \le 3,
\]
then the mean residence time
\(
\tau_{\mathrm{off}}
\)
remains finite while
\(
\langle a^2\rangle_{\mathrm{off}}
\)
diverges, implying that
\(
\langle x^2\rangle_{\mathrm{off}}
\)
also diverges. Such heavy tails, however, do not affect the injected power, which remains finite and 
depends only on the mean residence times [Eq.~(\ref{eq:dW-0})].
The divergence of
\(
\langle x^2\rangle_{\rm off}
\)
reflects the fact that the stationary released-state distribution inherits a power-law tail from the waiting-time statistics.



For exponential waiting-time distributions, for which 
\(\langle x^2\rangle_{\rm off}=\langle x^2\rangle_{b,\rm on}\) and
\(\langle x^2\rangle_{\rm on}=\langle x^2\rangle_{b,\rm off}\),
Eq.~(\ref{eq:x2-SR}) recovers the previously reported constant-rate results
\cite{JOP-Santra-2021,JOPA-Alston-2022,PRE-Frydel-2024},
\ba
\langle x^2\rangle_{\rm off}
&=&
\frac{k_BT}{K}
\left(
\frac{\tau_{\rm off}+\tau_{\rm on}}{\tau_{\rm on}}
\right)
+
2D\tau_{\rm off},
\nonumber\\
\langle x^2\rangle_{\rm on}
&=&
\frac{k_BT}{K}
\left(
\frac{\tau_{\rm off}+\tau_{\rm on}}{\tau_{\rm on}}
\right).
\ea
Also, in the limit of stochastic resetting with instantaneous returns,
corresponding in the present model to \(K\to\infty\),
the confined state collapses to a point at the origin and
\(
\langle x^2\rangle_{\rm on}\to0
\).
The only nontrivial stationary moment is then
\be
\langle x^2\rangle_{\rm off}
=
\frac{D}{\tau_{\rm off}}
\langle a^2\rangle_{\rm off},
\ee
which coincides with the standard result for diffusion with stochastic resetting.

\section{ Variance-space formulation}
\label{sec:gaussian-superposition}

For a harmonic potential with either time-dependent stiffness or time-dependent temperature, an initially Gaussian 
distribution remains Gaussian at all times~\cite{PRE-Frydel-2024}.  
This property follows from the linearity of the drift term in the corresponding Fokker--Planck equation.
This perspective simplifies the dynamics, since the problem no longer requires generating stochastic trajectories in $x$-space, where the dynamics explicitly contains diffusive noise.

For the confined state, the variance relaxes toward the equilibrium value
\(
\langle x^2\rangle \to D\tau_K
\),
where
\(
\tau_K = 1/(\mu K)
\).
Using the Langevin equation for the particle trajectory in the trapped state 
\(
d x  =  -\mu K x \, dt  +  \sqrt{2D} \, dB_t
\),
where \(B_t\) is standard Brownian motion, one obtains
\be
\frac{d}{dt}  \langle x^2\rangle = -\frac{2}{\tau_K}\langle x^2\rangle + 2D,
\label{eq:dx2-on}
\ee
whereas for the released state one finds
\be
\frac{d}{dt}\langle x^2\rangle = 2D.
\label{eq:dx2-off}
\ee
In both cases, the variance evolves deterministically because the diffusive noise has already been averaged over in the second moment dynamics.

To simplify things, we introduce the dimensionless variance
\[
s = \frac{\langle x^2\rangle}{D\tau_K} - 1
\]
which measures the excess variance relative to equilibrium and may be interpreted as an effective nonequilibrium temperature.
Using Eqs.~(\ref{eq:dx2-on}) and~(\ref{eq:dx2-off}), the dynamics of \(s\) in the released and confined states is
\be
\dot s = \frac{2}{\tau_K}, \qquad \dot s = -\frac{2}{\tau_K} s,
\label{eq:ds}
\ee
respectively.  We could alternatively write this as
\be
ds = \frac{2}{\tau_K} dt  -   \frac{2}{\tau_K} \sigma_t \, s \,dt
\label{eq:ds-langevin}
\ee
with $\sigma_t$ alternating between two values $\sigma_t\in \{0,1\}$  at times drawn from $r_{\rm off}$ and $r_{\rm on}$.

Equation~(\ref{eq:ds-langevin}) therefore defines a stochastic dynamics directly in variance space, generating the stationary distributions \(p_{\rm off}(s)\) and \(p_{\rm on}(s)\).
Both distributions are normalized, $\int_0^{\infty} ds\, p_i(s)=1$.  

The stationary spatial distributions are then reconstructed as
\be
n_i(x) = \int_0^\infty ds\,p_i(s)\,\rho_G(x;s),
\label{eq:nG}
\ee
where
\[
\rho_G(x;s) = \frac{1}{\sqrt{2\pi (s+1)D\tau_K}} \exp\!\left[ -\frac{x^2}{2(s+1)D\tau_K} \right]
\]
is a Gaussian distribution with dimensionless variance \(s\).

Eq.~(\ref{eq:nG}) shows that the fluctuating trap generates an ensemble of equilibrium-like Gaussian states characterized by different effective temperatures. 
The spatial distribution thus becomes a superposition of equilibrium Gaussian states, while all nonequilibrium information is encoded in the variance distribution \(p_i(s)\).
The lower limit \(s=0\) in that equation corresponds to the equilibrium variance of the confined harmonic trap. 
During the released phase, diffusion increases the variance, so \(s\ge 0\).

Since the dynamics in \(s\)-space is deterministic between switching events, stationarity requires local conservation of probability current,
\(
j(s)=0,
\)
where
\[
j(s) = \frac{\tau_{\rm off}}{\tau_{\rm off}+\tau_{\rm on}} \frac{2}{\tau_K}\,p_{\rm off}(s) - \frac{\tau_{\rm on}}{\tau_{\rm off}+\tau_{\rm on}} \frac{2}{\tau_K} s \,p_{\rm on}(s).
\]
Here
\(
\tau_i/(\tau_{\rm off}+\tau_{\rm on})
\)
is the stationary occupation probability of state \(i\), and the local velocities are given in Eq.~(\ref{eq:ds}).  
The zero-current condition yields
\be
p_{\rm on}(s) = \frac{\tau_{\rm off}}{\tau_{\rm on}} \frac{p_{\rm off}(s)}{s}.
\label{eq:pon-poff-universal}
\ee
Equation~(\ref{eq:pon-poff-universal}) is a general relation between the two variance distributions and is independent of the waiting-time distributions \(r_i\).
It shows that confinement universally suppresses large-variance fluctuations by one additional inverse power of \(s\).

\subsection{Stationary formalism}

Using the variance-space dynamics in Eq.~(\ref{eq:ds-langevin}), we next derive a stationary formalism for the distributions \(p_i(s)\). 
The structure is analogous to the stationary formalism previously obtained for the spatial distributions in Eqs.~(\ref{eq:b1-sol-2B}) and~(\ref{eq:pi-sol-2B}).
For the stationary variance distributions, one obtains 
\ba
&& p_{\rm off}(s) = \frac{1}{\tau_{\rm off}} \int_0^\infty ds'\,p_{b,\rm off}(s') \int_0^\infty da\,S_{\rm off}(a)\, P_{\rm off}( s,s',a), \nonumber\\ 
&& p_{\rm on}(s) = \frac{1}{\tau_{\rm on}} \int_0^\infty ds'\,p_{b,\rm on}(s') \int_0^\infty da\,S_{\rm on}(a)\, P_{\rm on}( s,s',a), \nonumber\\ 
\ea
while the corresponding birth distributions satisfy
\ba
&& p_{b,\rm off}(s)   =   \int_0^\infty ds'\,p_{b,\rm on}(s') \int_0^\infty da\,r_{\rm on}(a)\, P_{\rm on}( s,s',a), \nonumber\\ 
&& p_{b,\rm on}(s)   =   \int_0^\infty ds'\,p_{b,\rm off}(s') \int_0^\infty da\,r_{\rm off}(a)\, P_{\rm off}( s,s',a). \nonumber\\ 
\ea
The key simplification is that the propagators in variance space are deterministic
and therefore reduce to delta functions corresponding to the dynamics in Eq.~(\ref{eq:ds}),
\ba
&& P_{\rm off}( s,s',a)  =  \delta\!\left( s - s' - \frac{2a}{\tau_K} \right) \nonumber\\ 
&& P_{\rm on}( s,s',a)  =  \delta\!\bigg( s  - s'  e^{-2a/\tau_K} \bigg).  
\ea
Integrating over the delta functions yields 
\ba
p_{\rm off}(s) &=& \frac{\tau_K}{2\tau_{\rm off}} \int_0^s ds'\, p_{b,\rm off}(s')\, S_{\rm off}\!\left[ \frac{\tau_K}{2}(s-s') \right], \nonumber\\
p_{\rm on}(s) &=& \frac{\tau_K}{2\tau_{\rm on}s} \int_s^{\infty} ds'\, p_{b,\rm on}(s')\, S_{\rm on}\!\left[ \frac{\tau_K}{2} \ln\!\left( \frac{s'}{s} \right) \right],
\label{eq:ps-integral}
\ea
and 
\ba
p_{b,\rm on}(s) &=& \frac{\tau_K}{2} \int_0^s ds'\, p_{b,\rm off}(s')\, r_{\rm off}\!\left[ \frac{\tau_K}{2}(s-s') \right], \nonumber\\
p_{b,\rm off}(s) &=& \frac{\tau_K}{2s} \int_s^\infty ds'\, p_{b,\rm on}(s')\, r_{\rm on}\!\left[ \frac{\tau_K}{2} \ln\!\left( \frac{s'}{s} \right) \right].
\label{eq:pbs-integral}
\ea
Because of the exact relation in Eq.~(\ref{eq:pon-poff-universal}), the second equation in Eq.~(\ref{eq:ps-integral}) is redundant, since \(p_{\rm on}\) follows directly from \(p_{\rm off}\).

\subsection{Exact solutions}

Obtaining exact solutions of Eq.~(\ref{eq:ps-integral}) and Eq.~(\ref{eq:pbs-integral}) for general waiting-time distributions \(r_i\) is challenging. 
In this section, we consider two specific solvable cases: memoryless and deterministic switching. 
For waiting-time distributions with algebraic tails, only asymptotic analysis can be carried out.

For memoryless switching, we have
\(
\tau_i r_i = S_i.  
\)
Using Eq.~(\ref{eq:pon-poff-universal}), one obtains
$$
p_{\rm off}(s)  =  c_{\rm on} \int_0^s ds'\, \frac{p_{\rm off}(s')}{s'} e^{-c_{\rm off}(s-s')}, 
$$
where to simplify expressions, we define 
$$
c_{\rm off} =  \frac{\tau_K}{2\tau_{\rm off}}, \quad c_{\rm on} =  \frac{\tau_K}{2\tau_{\rm on}}.  
$$
The solution of this integral equation is a Gamma distribution,
\be
p_{\rm off}(s)
=
\frac{
c_{\rm off}^{\,c_{\rm on}+1}
}{
\Gamma(c_{\rm on}+1)
}
s^{c_{\rm on}}
e^{-c_{\rm off}s}.
\ee
The corresponding distribution \(p_{\rm on}(s)\) follows directly from Eq.~(\ref{eq:pon-poff-universal}).

Another exactly solvable case is deterministic switching. 
In this case, the birth distributions in variance space collapse to delta functions, since every cycle begins at the same value of \(s\). 
During the released phase, however, \(s\) grows linearly in time. 
Consequently, the stationary distribution \(p_{\rm off}(s)\) becomes uniform on a finite interval,
\be
p_{\rm off}(s) = c_{\rm off},
\ee
on the interval
\[
\frac{1}{c_{\rm off}} \frac{e^{-1/c_{\rm on}}}{1-e^{-1/c_{\rm on}}} \le s \le \frac{1}{c_{\rm off}} \frac{1}{1-e^{-1/c_{\rm on}}}.  
\]

We can also look into the asymptotic behavior of distributions with algebraic tails,
\[
r_{\rm off}(a)\sim C a^{-k},
\qquad
k>2,
\]
so that the corresponding survival function behaves as
\[
S_{\rm off}(a)\sim  \frac{C}{k-1} a^{1-k}.
\]
Note that only algebraic tails in the released state lead to algebraic spatial distributions.
Algebraic tails in \(r_{\rm on}\) do not generate algebraic spatial tails because confinement exponentially suppresses large variances.

To obtain an asymptotic relation for \(p_{\rm off}\), we start from the integral relation
\[
p_{\rm off}(s) = \frac{\tau_K}{2\tau_{\rm off}} \int_0^s ds'\, p_{b,\rm off}(s') S_{\rm off}\!\left[ \frac{\tau_K}{2}(s-s') \right].
\]
For large \(s\), the birth distribution remains concentrated at finite \(s'\), while the survival function varies on the scale of \(s\). 
The dominant contribution therefore comes from the regime \(s'\ll s\), so that, for the purpose of extracting the leading tail, one may treat \(p_{b,\rm off}\) as a point source, 
\(
p_{b,\rm off}(s')\to\delta(s').
\)
This leads to
\be
p_{\rm off}(s) \sim \frac{\tau_K}{2\tau_{\rm off}} S_{\rm off}\!\left( \frac{\tau_K}{2}s \right) \sim \frac{C}{k-1} \frac{1}{\tau_{\rm off}} \left(\frac{2}{\tau_K}\right)^{k-2} s^{1-k}.
\label{eq:p-tail}
\ee
The asymptotic behavior of the corresponding spatial distributions \(n_i(x)\) follows from Eq.~(\ref{eq:nG}), 
\be
n_{\rm off}(x) \sim \frac{C}{k-1} \frac{ \Gamma\!\left(k-\frac32\right) }{ \sqrt{\pi}\,\tau_{\rm off} } (4D)^{k-2}  |x|^{3-2k}.
\label{eq:n-off-tail}
\ee
At the level of exponents, this gives the sequence
\[
r_{\rm off}(a)\sim a^{-k}
\quad\Rightarrow\quad
p_{\rm off}(s)\sim s^{1-k}
\quad\Rightarrow\quad
n_{\rm off}(x)\sim |x|^{3-2k}.
\]

For the confined state, the corresponding asymptotic behaviors are
\be
p_{\rm on}(s)  \sim \frac{C}{k-1} \frac{1}{\tau_{\rm on}} \left(\frac{2}{\tau_K}\right)^{k-2} s^{-k}.
\label{eq:p-on-tail}
\ee
where we used Eq.~(\ref{eq:pon-poff-universal}), and
\be
n_{\rm on}(x) \sim \frac{C}{k-1} \frac{\Gamma\!\left(k-\frac12\right)}{\sqrt{\pi}\,\tau_{\rm on}} \frac{\tau_K}{2} (4D)^{k-1} |x|^{1-2k}, 
\label{eq:n-on-tail}
\ee
which at the exponent level yields the sequence
\[
r_{\rm off}(a)\sim a^{-k}
\quad\Rightarrow\quad
p_{\rm on}(s)\sim s^{-k}
\quad\Rightarrow\quad
n_{\rm on}(x)\sim |x|^{1-2k}.
\]

\begin{figure}[t]
\begin{center}
\includegraphics[width=0.25\textwidth]{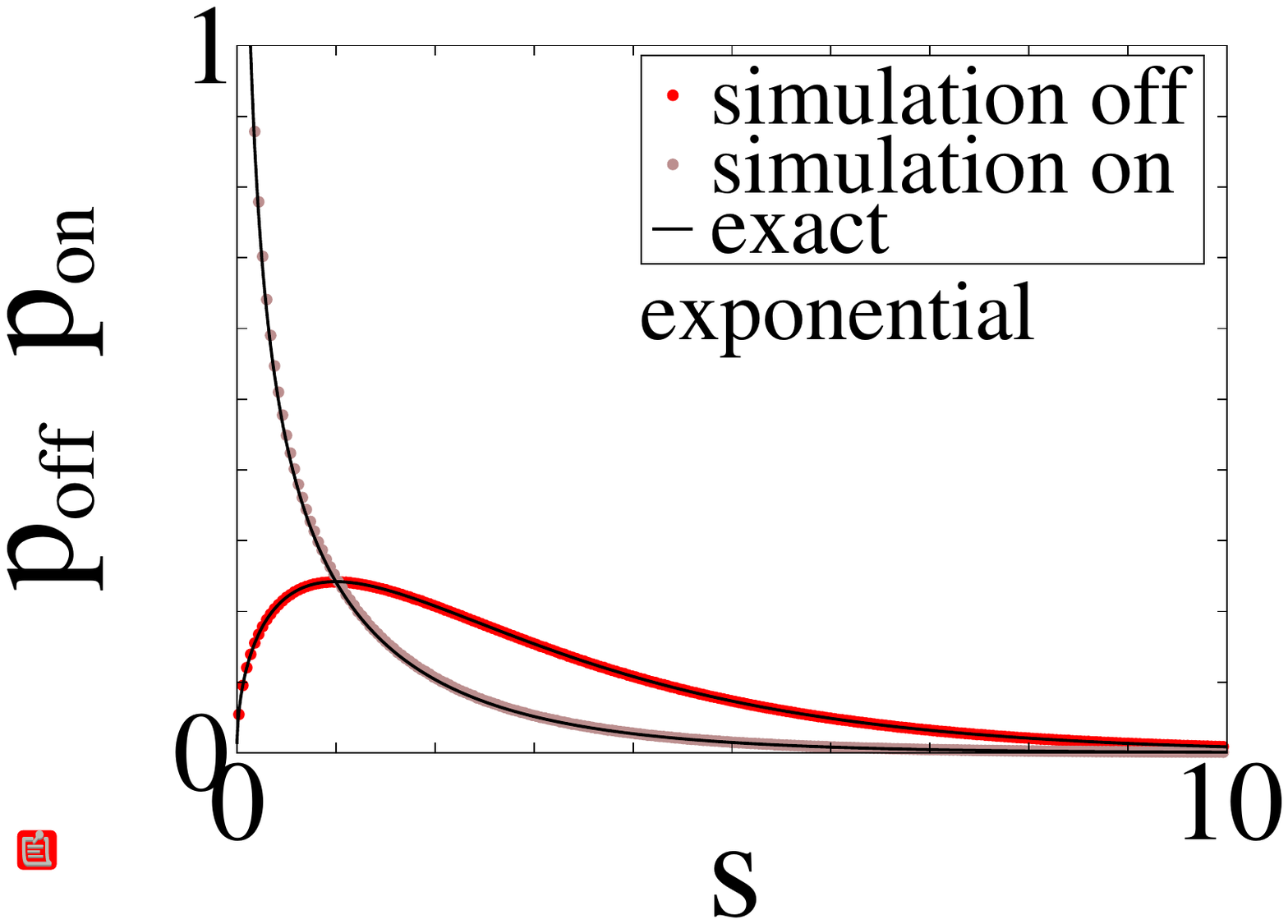}\\[0.2cm]
\includegraphics[width=0.25\textwidth]{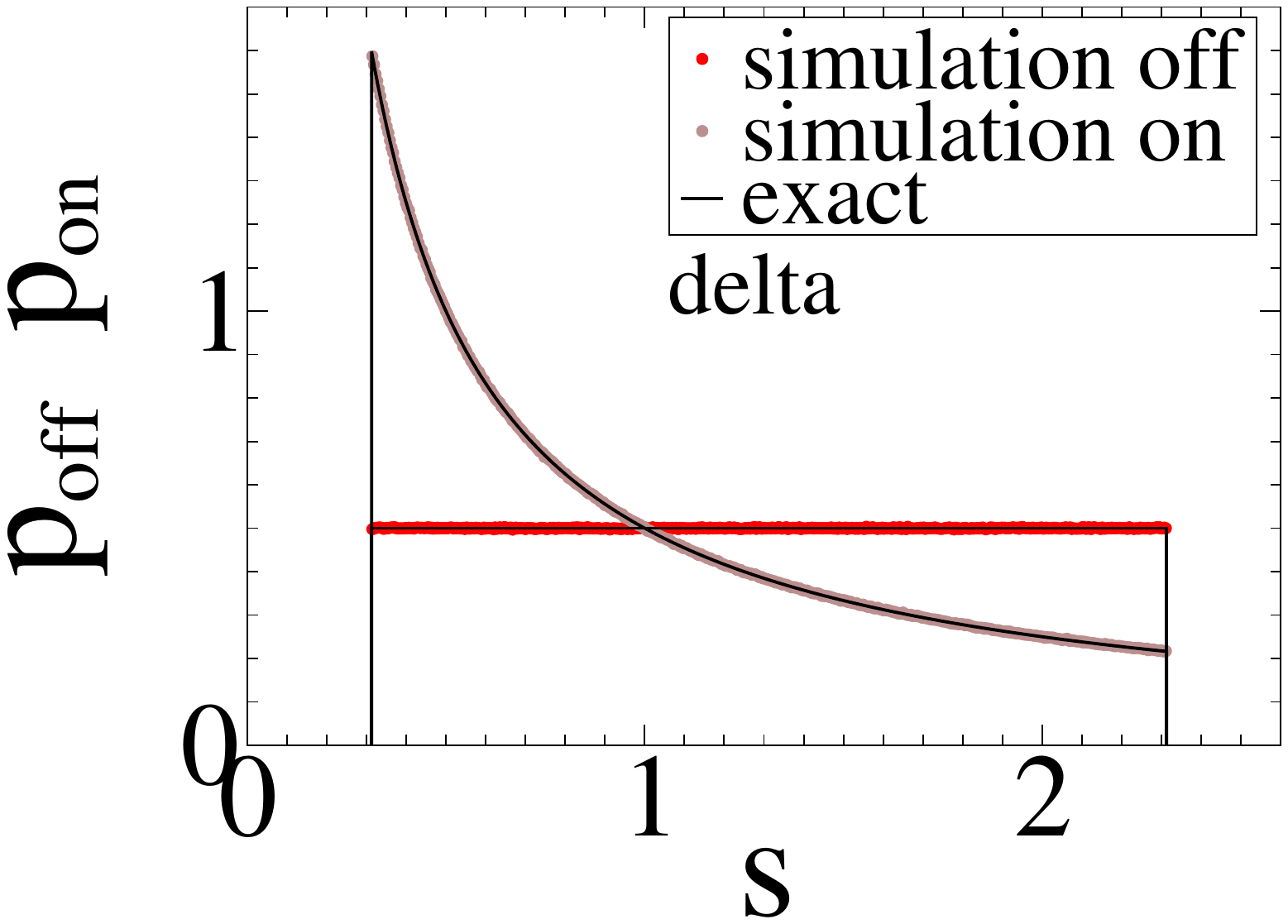}\\[0.2cm]
\includegraphics[width=0.25\textwidth]{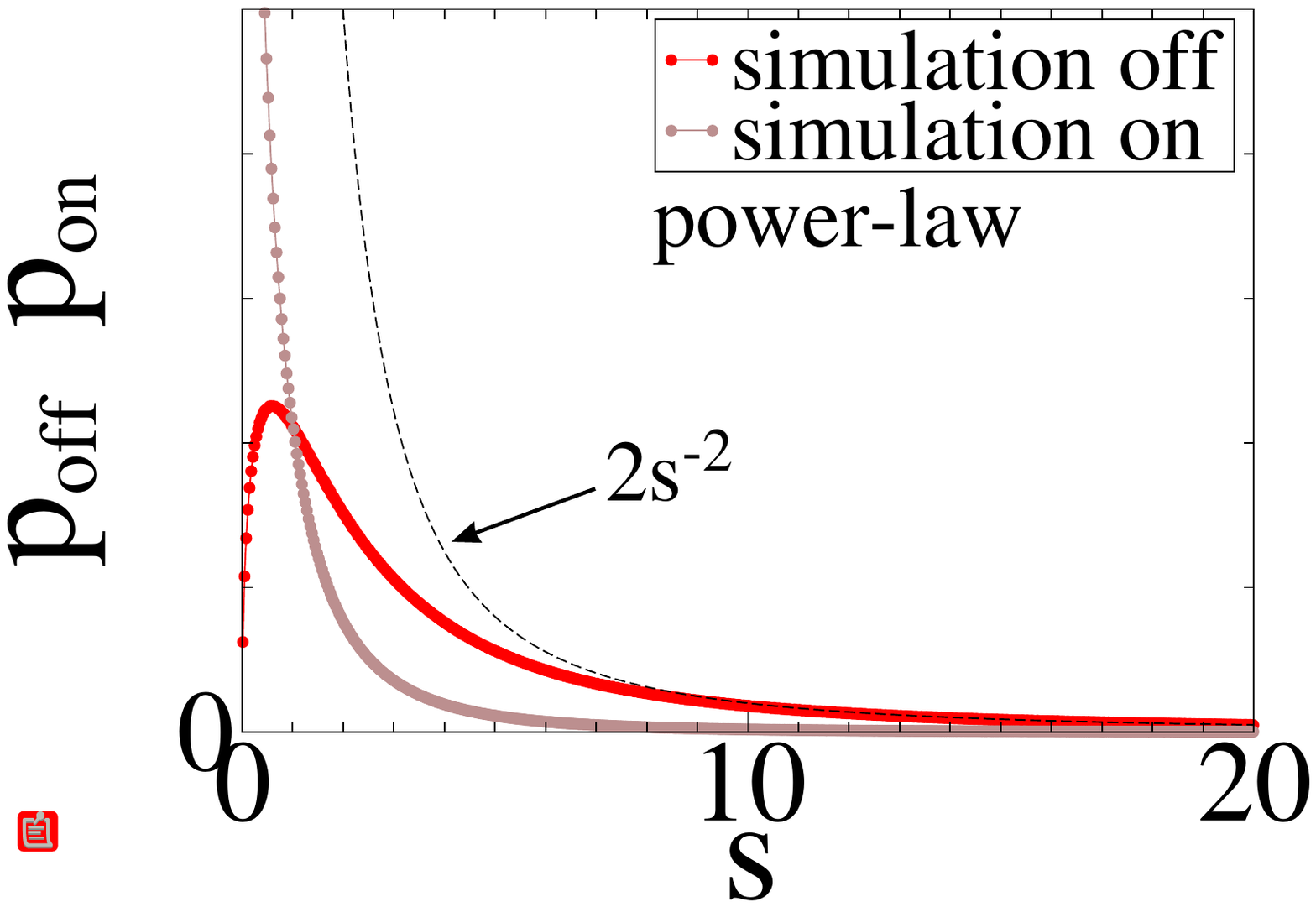}
\end{center}
\caption{
Variance-space distributions \(p_i(s)\) for exponential, deterministic, and algebraic waiting-time distributions.
Parameters: \(\tau_K=\tau_{\rm off}=\tau_{\rm on}=k_BT=1\).
For the algebraic case, we use \(r_{\rm off}\) from Eq.~(\ref{eq:powerlaw-ri}) with \(k=3\), corresponding to \(C=2\) 
and therefore \(p_{\rm off}(s)\sim 2s^{-2}\).
}
\label{fig:p}
\end{figure}

In Fig.~(\ref{fig:p}) we plot the variance distributions \(p_i(s)\) for different switching statistics and compare the analytical predictions with numerical simulations.
\graphicspath{{figures/}}
For the algebraic distribution \(r_{\rm off}\), the numerical results for \(p_{\rm off}(s)\) agree with the asymptotic prediction in Eq.~(\ref{eq:p-tail}).

\begin{figure}[t]
\begin{center}
\includegraphics[width=0.25\textwidth]{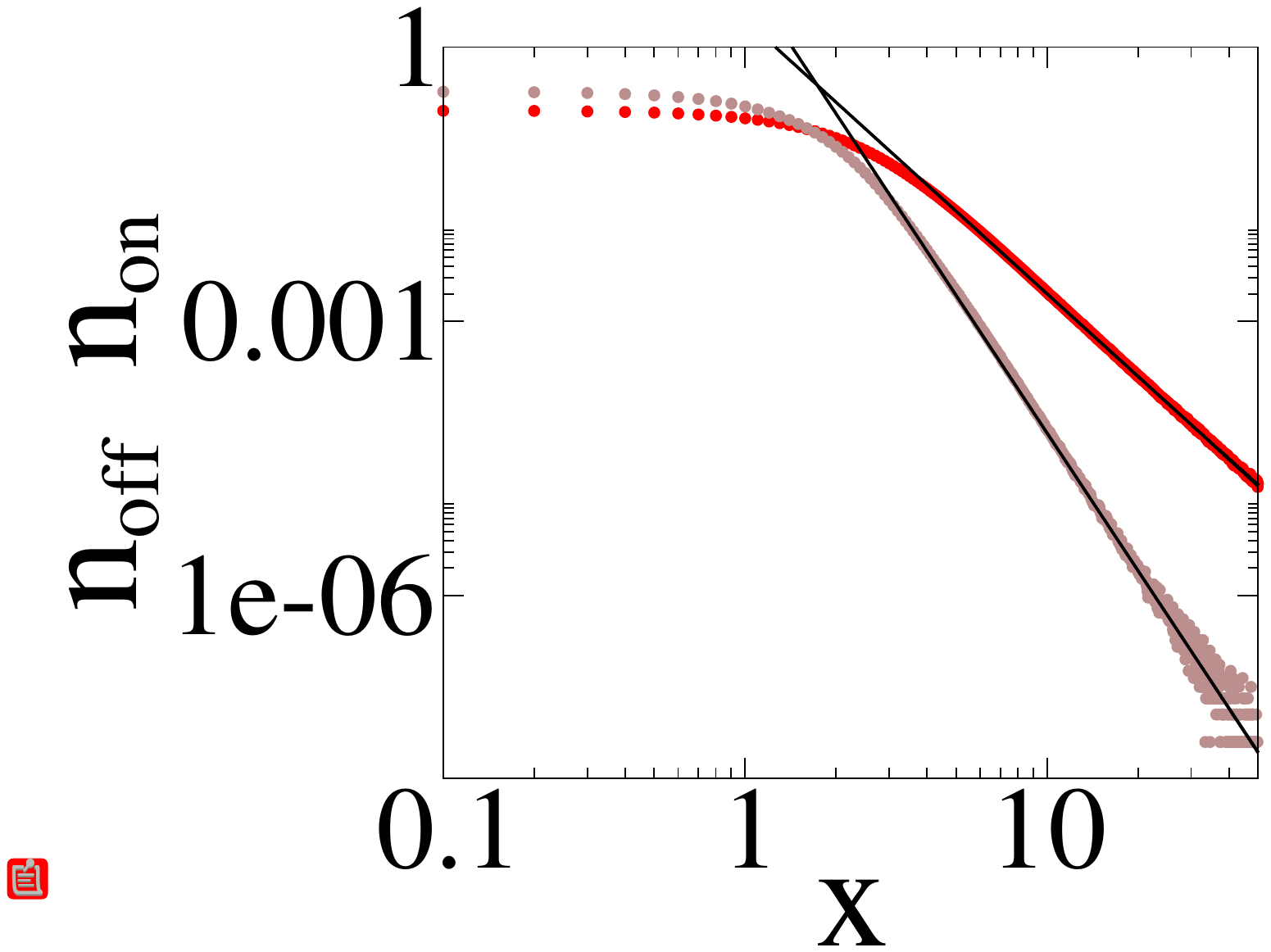}\\[0.2cm]
\end{center}
\caption{
Spatial distributions \(n_i(x)\) shown on a log-log scale for algebraic waiting-time distributions corresponding to Eq.~(\ref{eq:powerlaw-ri}) with \(k=3\).
Parameters: \(\tau_K=\tau_{\rm off}=\tau_{\rm on}=D=1\), for which \(n_{\rm off}(x)\sim 2|x|^{-3}\) and \(n_{\rm on}(x)\sim 6|x|^{-5}\).
}
\label{fig:n}
\end{figure}

In Fig.~(\ref{fig:n}) we show the spatial distributions \(n_i(x)\) on a log-log scale for \(k=3\) and compare their asymptotic tails with the predictions 
in Eq.~(\ref{eq:n-off-tail}) and Eq.~(\ref{eq:n-on-tail}).
The agreement with simulations confirms the accuracy of the asymptotic predictions.
\graphicspath{{figures/}}

\subsection{Moments}

Although the integral equations in Eqs.~(\ref{eq:ps-integral}) and~(\ref{eq:pbs-integral}) are generally difficult to solve explicitly for arbitrary 
waiting-time distributions $r_i$, analytical results can still be obtained for the moments.
To obtain such relations, we start by 
multiplying Eq.~(\ref{eq:pbs-integral}) by \(s^m\) and integrating over \(s\).  This produces the iterative relation 
between different moments 
\be
\langle s^m\rangle_{b,\mathrm{off}} = R_m \sum_{\ell=0}^{m-1} \binom{m}{\ell} \alpha_{m-\ell} \langle s^\ell\rangle_{b,\mathrm{off}}, \qquad m\ge1, 
\ee
where the coefficients are defined as 
\[
\alpha_n = \left(\frac{2}{\tau_K}\right)^n \langle a^n\rangle_{r_{\rm off}},  \quad \langle a^n\rangle_{r_{\rm off}} = \int_0^\infty da\,a^n r_{\rm off}(a),
\]
and
\be
R_n = \frac{ \tilde r_{\rm on}\!\left(2n/\tau_K\right)}{1-\tilde r_{\rm on}\!\left(2n/\tau_K\right)}.
\ee
Here \(\tilde r_{\rm on}\) denotes the Laplace transform of \(r_{\rm on}\).

Applying the same operation to Eq.~(\ref{eq:ps-integral}) gives the stationary moments in the released state,
\be
\langle s^m\rangle_{\rm off} = \sum_{\ell=0}^{m} \binom{m}{\ell} \frac{\alpha_{m-\ell+1}} {(m-\ell+1)\alpha_1} \langle s^\ell\rangle_{b,\mathrm{off}}, 
\ee
and the moments in the confined state follow from Eq.~(\ref{eq:pon-poff-universal}),
\be
\langle s^m\rangle_{\rm on} = \frac{\tau_{\rm off}}{\tau_{\rm on}} \langle s^{m-1}\rangle_{\rm off}, \qquad m\ge1.
\ee
Finally, using the Gaussian superposition in Eq.~(\ref{eq:nG}), the even spatial moments, which we represent as 
\ba
&&\frac{\langle x^{2m}\rangle_{\rm off}} {\langle x^{2m}\rangle_{\rm eq}} - 1  = \sum_{\ell=1}^{m} \binom{m}{\ell} \langle s^\ell\rangle_{\rm off}, \nonumber\\ 
&&\frac{\langle x^{2m}\rangle_{\rm on}} {\langle x^{2m}\rangle_{\rm eq}} - 1 =     \frac{\tau_{\rm off}}{\tau_{\rm on}} \sum_{\ell=1}^{m} \binom{m}{\ell} \langle s^{\ell-1}\rangle_{\rm off}. \nonumber\\
\ea
where $\langle x^{2m}\rangle_{\rm eq}$ is the moment for an equilibrium distribution for a trap with static stiffness $K$.  
The formulas above represent the excess of the spatial moments relative to thermal equilibrium.

For algebraic waiting-time distributions \(r_{\rm off}(a)\sim a^{-k}\) we find 
\[
\langle x^{2m}\rangle_{\rm off}<\infty \quad \text{for } m <  k - 2,
\]
whereas
\[
\langle x^{2m}\rangle_{\rm on}<\infty \quad \text{for } m <  k - 1.
\]
The confined state therefore suppresses fluctuations sufficiently to shift the divergence hierarchy by one moment order.

\section{Conclusion}
\label{sec:conclusion}

We developed an age-structured formulation for a Brownian particle in a two-state fluctuating harmonic trap with arbitrary waiting-time distributions. By augmenting the state space with the age variable, the dynamics become Markovian in the enlarged state space and admit a local-in-time Fokker--Planck description. Integrating out the age variable yields a reduced evolution in which memory is encoded through the birth flux associated with switching events, rather than through explicit temporally nonlocal memory kernels.

Under steady-state conditions, the formalism leads to a closed set of self-consistent integral equations for the birth distributions and stationary spatial densities. Without solving for the full stationary distributions explicitly, we derived exact expressions for the second moments and showed that the waiting-time statistics enter through Laplace-transform quantities characterizing the switching process.

A central result is the distinction between energetic and spatial observables. While the average potential energy remains fixed at its equilibrium value independently of the switching statistics, as a consequence of a generalized virial relation valid for homogeneous potentials, the injected power and spatial fluctuations retain strong sensitivity to temporal memory. In particular, deterministic switching maximizes the injected power at fixed mean residence times, whereas increasingly intermittent protocols suppress dissipation.

In the stochastic-resetting limit, corresponding to alternating diffusion and confinement, the injected power becomes universal and depends only on the mean residence times, while spatial fluctuations remain sensitive to higher moments and asymptotic tails of the waiting-time distributions. 
To analyze these fluctuations, we introduced a variance-space formulation in which the stochastic dynamics in $x$-space is replaced by deterministic dynamics in the 
variance-space, interrupted by semi-Markovian switching. In this representation, the stationary state becomes a superposition of Gaussian states with fluctuating variance.

The variance-space formalism yields integral equations for the variance distributions, exact solutions for exponential and deterministic switching, asymptotic spatial tails for algebraic waiting-time distributions, and exact moment hierarchies. In particular, algebraic residence-time distributions in the released state generate algebraic tails in both variance and spatial distributions. The confined state exhibits systematically faster decay, leading to a shift in the divergence hierarchy of spatial moments by one moment order.

More broadly, the age-structured viewpoint developed here provides a transparent alternative to memory-kernel formulations for semi-Markovian nonequilibrium systems. Because the framework isolates the origin of memory in the residence-time statistics themselves, it should be applicable to a broader class of stochastic switching processes involving non-exponential dynamics, including active-matter and transport problems with internal-state switching.

\begin{acknowledgments}
D.F. acknowledges financial support from FONDECYT through grant number 1241694.  
\end{acknowledgments}

\section{Data availability}
The data that support the findings of this study are available from the corresponding author upon 
reasonable request.


\bibliography{general}


\end{document}